\documentclass[aps,pra,twocolumn,superscriptaddress]{revtex4-1}

\usepackage{graphicx}
\usepackage[rightcaption]{sidecap}
\usepackage{floatrow}
\usepackage{amsmath}
\usepackage{array}
\usepackage{color}
\newcolumntype{@}{>{\global\let\currentrowstyle\relax}}
\newcolumntype{^}{>{\currentrowstyle}}
\newcommand{\rowstyle}[1]{\gdef\currentrowstyle{#1}%
  #1\ignorespaces
}
\newcommand{\sOne}{$4s^2$ $^1S_0$}
\newcommand{\pTwo}{$4s4p$ $^3P_0$}
\newcommand{\pThree}{$4s4p$ $^3P_1$}
\newcommand{\pFour}{$4s4p$ $^3P_2$}
\newcommand{\dFive}{$3d4s$ $^3D_1$}
\newcommand{\dSix}{$3d4s$ $^3D_2$}

\newcommand{\dEight}{$3d4s$ $^1D_2$}
\newcommand{\pNine}{$4s4p$ $^1P_1$}
\newcommand{\sTen}{$4s5s$ $^3S_1$}
\newcommand{\pNineteen}{$4s5p$ $^1P_1$}
\newcommand{\fThirtyFour}{$3d4p$ $^1F_3$}
\newcommand{\pThirtySeven}{$4s6p$ $^1P_1$}
\newcommand{\fFortyTwo}{$4s4f$ $^1F_3$}
\newcommand{\pFiftyThree}{$4snp$ $^1P_1$}
\newcommand{\fFiftyNine}{$4s5f$ $^1F_3$}
\newcommand{\pSixtyNine}{$4s7p$ $^1P_1$}
\newcommand{\fSeventyFive}{$4s6f$ $^1F_3$}

\begin{document}

\title{Improved magneto-optical trapping of Ca}

\author{Michael Mills}
\author{Prateek Puri}
\affiliation{Department of Physics and Astronomy, University of California -- Los Angeles, Los Angeles, California, 90095, USA}
\author{Yanmei Yu}
\affiliation{Beijing National Laboratory for Condensed Matter Physics, Institute of Physics, Chinese Academy of Sciences, Beijing 100190,China}
\affiliation{Department of Physics, University of Nevada, Reno, Nevada 89557, USA}
\author{Andrei Derevianko}
\affiliation{Department of Physics, University of Nevada, Reno, Nevada 89557, USA}
\author{Christian Schneider}
\author{Eric R. Hudson}
\affiliation{Department of Physics and Astronomy, University of California -- Los Angeles, Los Angeles, California, 90095, USA}

\date{\today}

\begin{abstract}
We investigate the limiting factors in the standard implementation of the Ca magneto-optical trap. We find that intercombination transitions from the \pNineteen\ state used to repump electronic population from the \dEight\ state severely reduce the trap lifetime. We explore seven alternative repumping schemes theoretically and investigate five of them experimentally. We find all five of these schemes yield a significant increase in trap lifetime and consequently improve the number of atoms and peak atom density by as much as $\sim20\times$ and $\sim6\times$, respectively. One of these transitions, at 453 nm, is shown to approach the fundamental limit for a Ca magneto-optical trap with repumping only from the \dEight\ state, yielding a trap lifetime of $\sim$5~s.

\end{abstract}


\maketitle

\section{\label{Introduction}Introduction}
The magneto-optical trap (MOT)~\cite{Raab1987} is an integral part of atomic and molecular physics, where it is the starting point for a variety of experiments including precision tests of fundamental physics~\cite{Hamilton2015}, studies of quantum many-body physics~\cite{Kaufman2016}, and production of ultracold molecules~\cite{Rellergert2013, Barry2014}. At present, atomic MOTs have been constructed for atoms within Groups 1, 2, 6, 12, and 18, as well as the lanthanides. Extension to atoms in other Groups is often limited by the availability of appropriate laser technology for driving the necessary cooling transitions and complications due to the electronic structure of the atom. For example, if there are multiple electronic states below the upper electronic state of the primary laser cooling transition, then radiative decay into these lower levels can severely reduce, and even eliminate, the laser cooling force. For these reasons, the Group 1 atoms, with their lone optically active, unpaired electron, provide the simplest, and often best performing, MOTs.

Nonetheless, the same `complications' that can limit the laser cooling process often host interesting and useful phenomena. A prime example of this are the $^3P$ states of Group 2(-like) atoms, which, while detrimental to the performance of a standard MOT, allow the construction of next-generation optical atomic clocks that outperform the Cesium standard~\cite{Ludlow15}. One such MOT of this type is the Ca MOT. Calcium MOTs have been utilized in atomic optical clock experiments using the 657 nm $ ^3P_1~\leftarrow~^1S_0 $ intercombination line~\citep{Sterr04,Riehle05,NIST06} and have significant appeal due to their simplicity of construction as portable optical frequency standards~\cite{Amar}. However, despite this appeal, the details of the Ca electronic structure lead to relatively poor performance of Ca MOTs, including a short trap lifetime limited by optical pumping into dark states and a low achievable peak atomic density. For these reasons, other Group 2(-like) atoms such as Sr, Yb, and Hg have become more popular choices for optical frequency standards \cite{Ludlow15,DerKat11}.

Given the potential of Ca as a portable frequency standard, as well as its utility in our own experiments as a sympathetic coolant for molecular ions~\cite{Rellergert2013}, we have performed a detailed combined experimental and theoretical study of Ca MOT operation. Specifically,  relativistic many-body calculations are performed for the first 75 energy levels of the Ca atom, providing reliable electronic structure and transition matrix elements for this multi-electron atom. The results of this calculation are incorporated into a rate equation model for the populations in the Ca atom, which is used to evaluate specific repumping schemes and identify seven promising transitions. In total, we experimentally investigate five alternative repumping schemes and find that all of them yield Ca MOTs with lifetimes and atom numbers improved by $\sim10\times$ over the traditional scheme described in Ref.~\citep{Gruenert2001}. The best of these schemes, which utilizes repumping to a highly configuration-mixed state with a 453 nm repumping laser, produces a Ca MOT with lifetime, number, and density improved over the standard MOT by $\sim25\times$, $\sim20\times$, and $\sim6\times$, respectively.

In the remainder of this paper, we first present the details of the relativistic many-body calculation of the Ca energy levels and the resulting rate equation model of the Ca populations. We then use this rate equation model to explain the poor performance of the traditional Ca MOT. From this work, we propose seven alternative MOT operation schemes and experimentally investigate five of them. We characterize the differences in these MOT operation schemes, reporting the achievable MOT lifetimes, density, and trapped atom numbers, as well as the necessary repumping laser frequencies.  We conclude with discussion of the ideal repumping scheme for Ca MOT operation and possible extension to other Group 2(-like) atoms.

\section{\label{Theory}Relativistic many-body calculations of atomic structure}

\begin{table*}[t!]
\caption{\label{Tab:ThExpComparision} Comparison of selected NIST recommended and theoretical CI+MBPT values. The recommended  values of $\left| \langle f ||D|| i \rangle \right|$ are derived from  transition rates compiled in the NIST atomic spectra database~\cite{NIST_ASD} as $\sqrt{1.5\times10^{-8}A_{fi}\lambda^3g_f/303.8}$, where $A_{fi}$ is transition rate in $10^8$/s, $\lambda$ is wavelength in $\AA$, $g_f=(2J_f+1)$ with $J_f$ being  the final state angular momentum.}
\begin{ruledtabular}
\begin{tabular}{@c ^c ^c ^c ^c ^c}
\multicolumn{2}{c}{States}   &  \multicolumn{2}{c}{$\Delta E$,  $\mathrm{cm}^{-1}$} &  \multicolumn{2}{c}{$\left| \langle f ||D|| i \rangle \right|$, a.u.} \\
 initial           &     final          & CI+MBPT   & NIST      & CI+MBPT & NIST \\ \hline
 $4s4p~^1\!P_1$    &  $4s^2~^1\!S_0$    &    23491  &    23652  &    4.9  &4.9   \\
 $4s5s~^3\!S_1$    &  $4s4p~^3\!P_0$    &    16604  &    16382  &    1.8  &1.8      \\
 $4s5s~^3\!S_1$    &  $4s4p~^3\!P_1$    &    16550  &    16329  &    3.1  &3.2     \\
 $4s5s~^3\!S_1$    &  $4s4p~^3\!P_2$    &    16440  &    16224  &    4.0  &4.1    \\
 $4s5p~^1\!P_1$    &  $3d4s~^1\!D_2$    &    14259  &    14882  &    2.4  &2.3      \\
 $4p^2~^1\!D_2$    &  $4s4p~^1\!P_1$    &    17691  &    17068  &    5.6  &5.7      \\
 $4s6p~^1\!P_1$    &  $4s^2~^1\!S_0$    &    41788  &    41679  &    0.4  &0.6  \\
 $4s4f~^1\!F_3$    &  $3d4s~^1\!D_2$    &    19943  &    20494  &    3.5  &2.8   \\
 $4s5f~^1\!F_3$    &  $3d4s~^1\!D_2$    &    22421  &    22955  &    2.3  &2.3    \\
 $4s7p~^1\!P_1$    &  $4s^2~^1\!S_0$    &    46975  &    45425  &    0.6  &0.5   \\
 $4snp~^1\!P_1$    &  $4s^2~^1\!S_0$    &    44383  &    43933  &    0.8  &0.7   \\	
\end{tabular}
\end{ruledtabular}
\end{table*}

The analysis of MOT performance requires estimates of electric-dipole transition rates between the 75 lowest-energy levels of Ca, including both spin-allowed and spin-forbidden (intercombination) transitions. While the energy levels are well established, transition rates (811 possible channels) are generally unknown, and we evaluated them
using methods of  relativistic many-body theory. {\em Ab initio} relativistic calculations are necessary as the analysis requires inclusion of transition amplitudes that are forbidden at the non-relativistic level.

Calcium is an atom with two valence electrons outside a tightly bound core. We employ a systematic formalism that combines advantages of both the configuration interaction (CI) method and many-body perturbation theory (MBPT), the CI+MBPT method~\cite{DzuFlaKoz96}. The CI+MBPT method has been used extensively for evaluation of atomic properties (see, e.g., review~\cite{DerPor11} for optical lattice clock applications and references therein).
Relativistic effects are included exactly, as the formalism starts from the Dirac equation and employs relativistic bi-spinor wave functions throughout the entire calculation.  In our treatment, the CI model space is limited to excitations of valence electrons. Contributions involving excitations of core electrons are treated within MBPT.  In this approach, we first solve for the valence electron orbitals and energies in the field of core electrons. The one-electron effective potential includes  both the frozen-core Dirac-Hartree-Fock (DHF $V^{N-2}$)  and self-energy (core-polarization) potentials. The self-energy correction is computed using second-order MBPT diagrams involving virtual core excitations. At the next step, the computed one-electron valence orbitals are used to diagonalize the atomic Hamiltonian in the model space of two valence electrons within the CI method.  The CI Hamiltonian includes the residual (beyond DHF) Coulomb interaction between the valence electrons and their core-polarization-mediated interaction. The latter was computed in the second-order MBPT. This step yields  two-electron wave-functions and energies. Finally, with the obtained wave-functions we calculated the required electric-dipole matrix elements. In calculations of transition rates we used experimental energy intervals and the computed CI+MBPT matrix elements.

We used two independent CI+MBPT implementations: (i) by the Reno group (see the discussion of the earlier version in Ref.~\cite{AD01}) and (ii) a recently published package~\cite{Kozlov2015CPC}. The practical goal of the calculations was not to reach the highest possible accuracy, but rather to generate the large amount of data needed for the transition array involving the 75 lowest-energy levels. An additional computational challenge was the inclusion of high angular momenta states, e.g.,  the $4s5g$ $^3G$ state.  The Reno code was run on a large basis set but without including  core-polarization-mediated interaction in the CI Hamiltonian due to considerable computational costs. The production runs with the package of Ref.~\cite{Kozlov2015CPC} employed a  smaller basis set (due to code limitations) but treated the correlation problem more fully. Our final values combine the outputs of the two codes. The bulk of the results comes from the package of Ref.~\cite{Kozlov2015CPC}. These results are augmented with the $ns8s$ states from the Reno code due to the limited number of roots in the package of Ref.~\cite{Kozlov2015CPC}.

We assessed the quality of the calculations by comparing the CI+MBPT energies with the NIST recommended values~\cite{NIST_ASD} and by comparing computed transition rates with the NIST recommended data, when available. A representative comparison is presented in Table~\ref{Tab:ThExpComparision}.    
From this table, we see that the energy intervals are reproduced with a few percent accuracy, while matrix elements with 1-30\% accuracy. Such accuracies are sufficient for the goals of this paper.
The full transition array with 811 entries together with a Mathematica front-end for accessing these data can be found in the supplementary materials. Detailed formatted tables will be published elsewhere.

\section{\label{RE}Rate equation model of C\MakeLowercase{a} electronic state populations}
Using these calculated transition rates, we create a rate model including the first 75 excited states of calcium. As an example, the differential equation for state $i$ with a laser driving from state $i$ to state $k$ is given by

\begin{multline*}
\frac{d}{dt}N_i = \sum_{j>i} A_{ji}N_j - \sum_{j<i} A_{ij} N_i\\ + A_{ki} \frac{\pi^2 c^3}{\hbar \omega_{ik}^3} \frac{I_{l}}{2\pi c} \frac{\Gamma_k}{(\omega_{ik}-\omega_{l})^2 + \frac{\Gamma_k^2}{4}} \left( N_k - \frac{2j_k + 1}{2j_i + 1}N_i \right) \,
\end{multline*}
where $N_i$ is the number of atoms in state $i$, $A_{ij}$ is the decay rate of state $i$ to $j$, $c$ is the speed of light in a vacuum, $\hbar$ is the reduced Planck constant, $\omega_{ik}$ is the angular transition frequency between state $i$ and $k$, $\omega_{l}$ ($I_{l}$) is the angular frequency (intensity) of the applied laser, $\Gamma_k$ is the natural linewidth of state $k$, and $j_i$ is the total angular momentum quantum number of state $i$.

\section{\label{StandardOperation}Evaluation of the standard C\MakeLowercase{a} MOT operation}
\begin{figure}[t]
\includegraphics[width=\columnwidth]{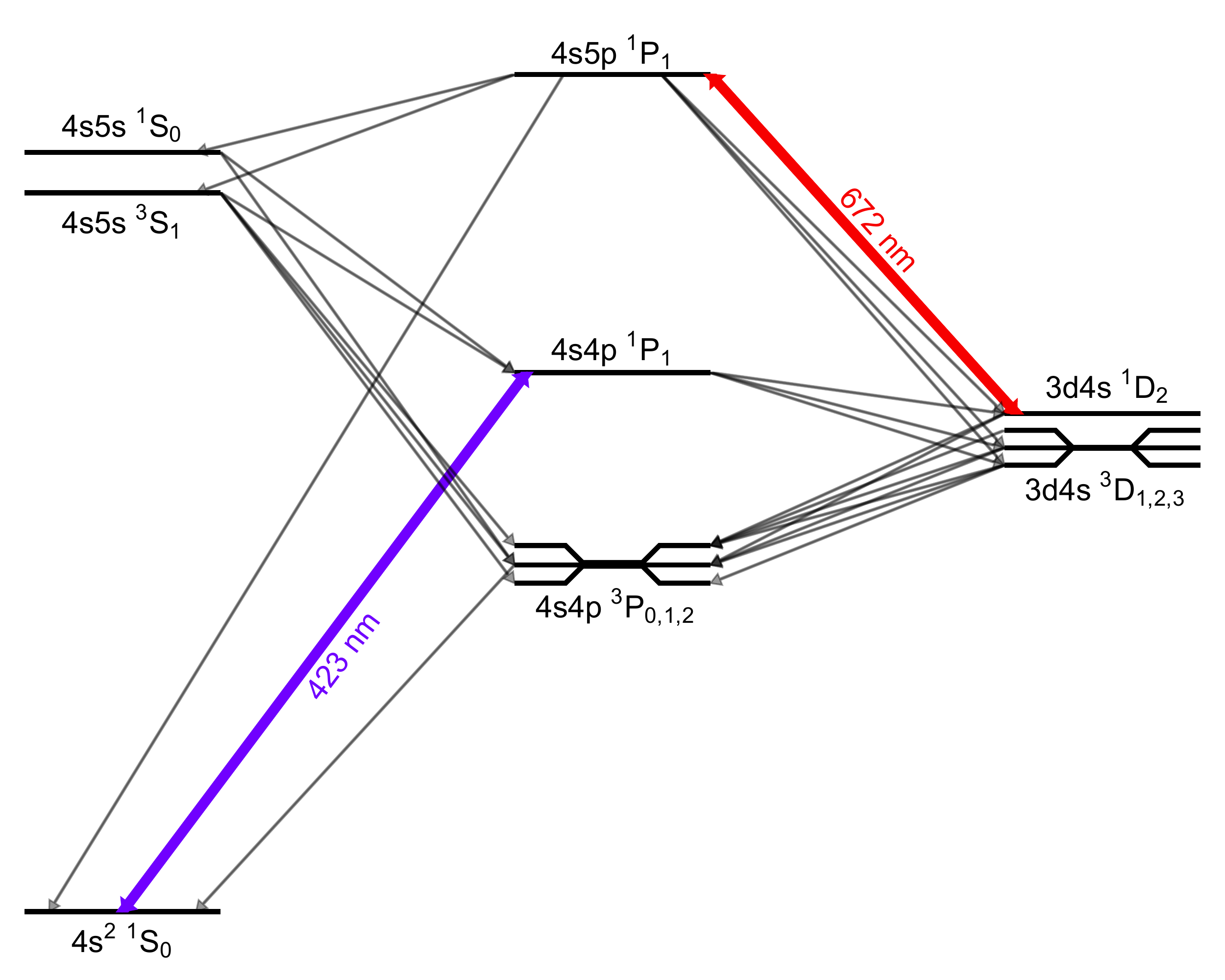}
\caption{\label{fig:levels672}Relevant level structure for operation of a standard calcium MOT. Laser cooling is accomplished on the 423 nm \pNine\ $\leftarrow$ \sOne\ transition. Atoms that decay to the \dEight\ state are repumped back into the cooling cycle via the 672 nm \pNineteen\ $\leftarrow$ \dEight\ transition, while those in the long-lived $4s4p~^3P_{0,2}$ states are lost from the MOT.}
\end{figure}

The standard implementation of a Ca MOT is formed by laser cooling on the strong \pNine $\leftarrow$ \sOne\ transition at 423~nm in the presence of an anti-Helmholtz magnetic field with gradient of 60 G/cm in the axial direction. This transition will incur loss from the laser cooling cycle primarily due to decay from the \pNine\ state to the \dEight\ state. This $^1D_2$ state, as shown in Fig. \ref{fig:levels672}, decays to the \pThree\ (83\% branching) and $^3P_2$ (17\% branching) states with a total lifetime of 1.71 ms \cite{Husain86}. The $^3P_1$ state decays to the ground state with a lifetime of 0.331 ms, while the $^3P_2$ state has a lifetime of 118 minutes, leading to loss from the laser cooling cycle \cite{Husain86,AD01}. This loss, which is proportional to the \pNine\ state population, limits the lifetime of the Ca MOT and according to the rate model with our MOT parameters leads to a MOT lifetime of 27 ms. As detailed later, we experimentally observe a MOT lifetime of 29(5)~ms in this configuration.

To extend the MOT lifetime, a repumping laser is usually added to drive the \pNineteen $\leftarrow$ \dEight\ transition at 672 nm in order to return electronic population in the \dEight\ level to the laser cooling cycle before it decays to the \pThree\ and $^3P_2$ states~\cite{Kurosu1992}. In this configuration, the rate equation model predicts that the MOT lifetime is increased to 98 ms for our MOT parameters.  As detailed later, we experimentally observe a MOT lifetime of 93(6)~ms in this configuration.

Interestingly, it is often assumed that the lack of a further increase in the MOT lifetime with this repumping scheme is an incomplete depletion of the \dEight\ state due to unfavorable branching ratios in the \pNineteen\ state~\cite{Kurosu1992}. This state decays primarily back to the \dEight\ state and only weakly back in the cooling cycle. However, the rate equation model reveals that the MOT lifetime is actually limited by the decay of the \pNineteen\ state to the \sTen , \dFive , and \dSix\ states, all of which decay primarily to the $4s4p$ $^3P_{0,1,2}$ states, as shown in Fig. \ref{fig:levels672}, as first pointed out in Ref.~\cite{Oates1999}. Specifically, according to the theoretical calculations, the \pNineteen\ state decays indirectly to the lossy \pTwo\ and $^3P_2$ states at a total rate of $8\times10^4$~s$^{-1}$, while the \dEight\ state decays to the \pFour\ state at a rate of only $80$~s$^{-1}$. With this understanding, the natural question arises: \textit{Is there an alternative repumping scheme that would suppress the loss into these triplet states?}

\section{\label{ImprovedOperation}Evaluation of alternative C\MakeLowercase{a} MOT operation schemes}

The ideal repumping laser out of the \dEight\ state would quickly transfer population from the $^1D_2$ state back into to the cooling cycle with perfect efficiency. With this idealized scheme, the rate model predicts a lifetime of 3.0 s with our MOT parameters. This lifetime is limited by the decay of the \pNine\ state to \dFive\ and $^3D_2$ states and is thus dependent on the \pNine\ state population; lowering the \pNine\ state population by decreasing 423 nm cooling laser intensity while maintaining reasonable MOT performance can extend the lifetime by $\sim2\times$.  Since this lifetime is similar to lifetimes set by other effects in most systems, such as collisions with background gas, it is likely unnecessary for the majority of applications to employ a more complicated multi-laser repumping scheme out of the $^3$P states like that used in Sr~\cite{Ludlow15}, especially since the longer lifetime of the \dEight\ and \pThree\  in Ca make this scheme less efficient.

\begin{figure}
\includegraphics[width=\columnwidth]{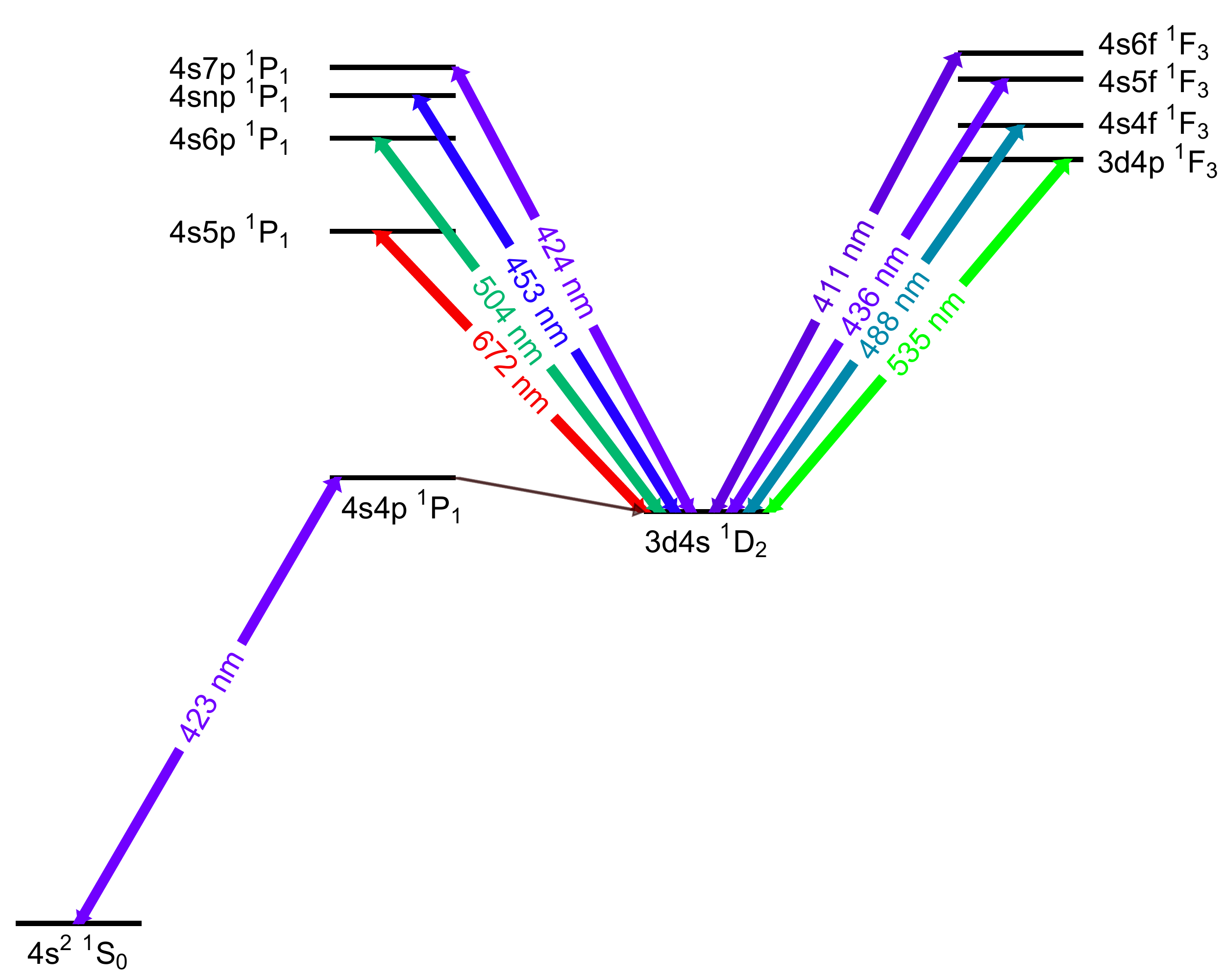}
\caption{\label{fig:levelsRepumps} Simplified calcium electronic level structure showing the eight repumping transitions considered here. All transitions except the 504~nm and 535~nm have been studied experimentally. The overall best Ca MOT performance is found when using the 453~nm \pFiftyThree\ $\leftarrow$ \dEight\ transition.}
\end{figure}

Therefore, for this work we chose to only explore single-laser repump transitions from the \dEight\ state with high branching ratios back into the laser cooling cycle. With this metric, we find that within the first 75 electronic states, there are seven reasonable alternative repumping transitions from the \dEight\ state, shown in Fig.~\ref{fig:levelsRepumps}, which go to states in the $^1P_1$ and $^1F_3$ manifolds. Using the rate equation model with our standard MOT parameters, we calculate the expected MOT lifetimes for these transitions, which are limited by optical pumping into the $^3P_{0,2}$ states, and present the results in Table~\ref{mainTable}.

Of these seven transitions, five are accessible by lasers available to us and we explore them using a standard six beam Ca MOT described in Ref.~\cite{Rellergert2013}. Briefly, in this system, laser cooling is provided by driving the \pNine\ $\leftarrow$ \sOne\ cooling transition with a total laser intensity of 63~mW/cm$^2$ detuned 34.4~MHz below resonance. The Ca MOT is loaded from an oven source placed $\sim$~3.5~cm away from the MOT. Atoms from the oven are decelerated by two `deceleration beams' with intensities 110~mW/cm$^2$ and 53~mW/cm$^2$ and detunings below resonance of 109~MHz and 318~MHz, respectively. The 672~nm traditional Ca MOT repump laser has an intensity of 11~mW/cm$^2$.

\begin{figure}
\centering

\begin{minipage}[t]{.49\columnwidth}
\centering
\includegraphics[width=\columnwidth]{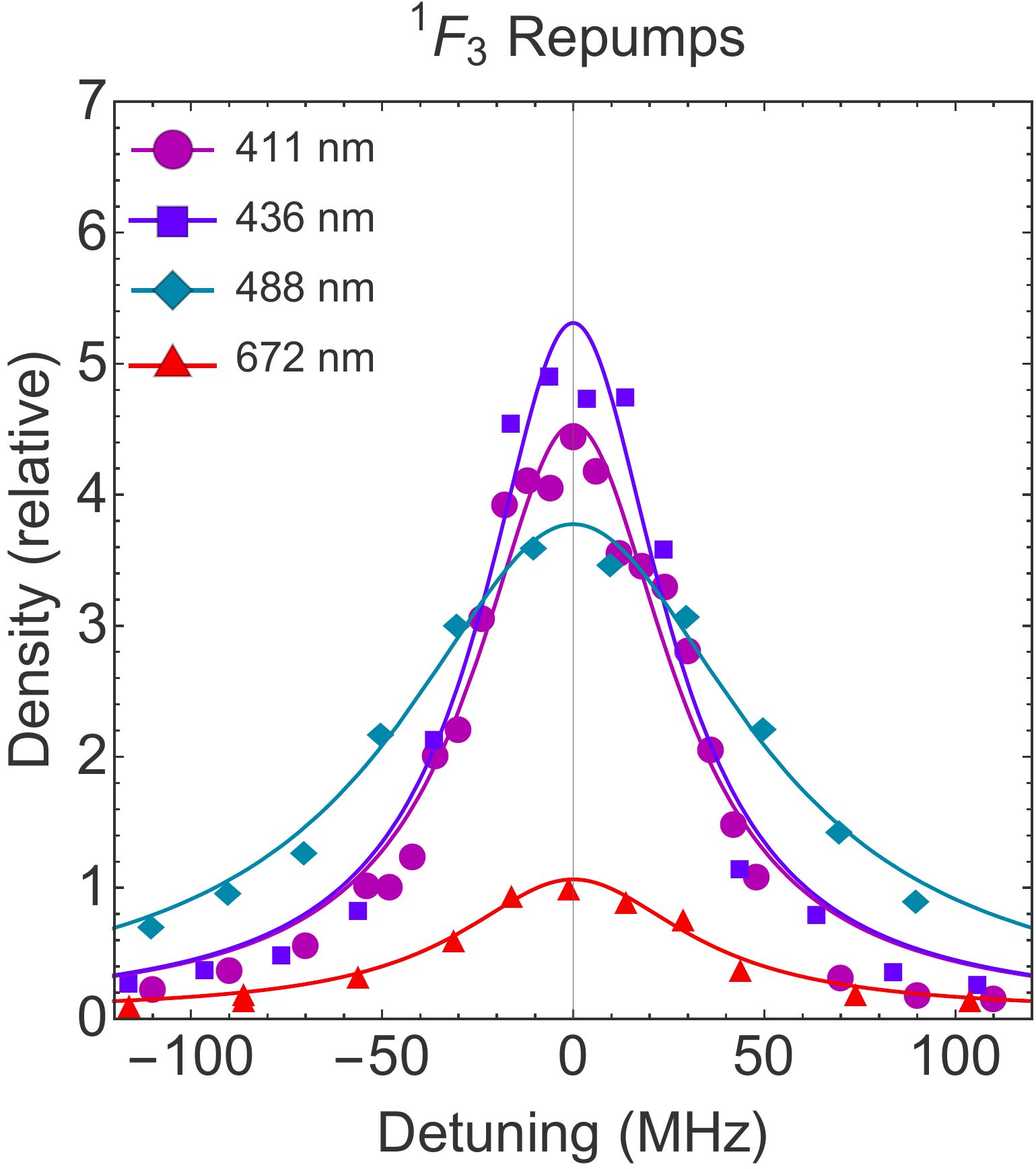}
\end{minipage}
\begin{minipage}[t]{.49\columnwidth}
\centering
\includegraphics[width=\columnwidth]{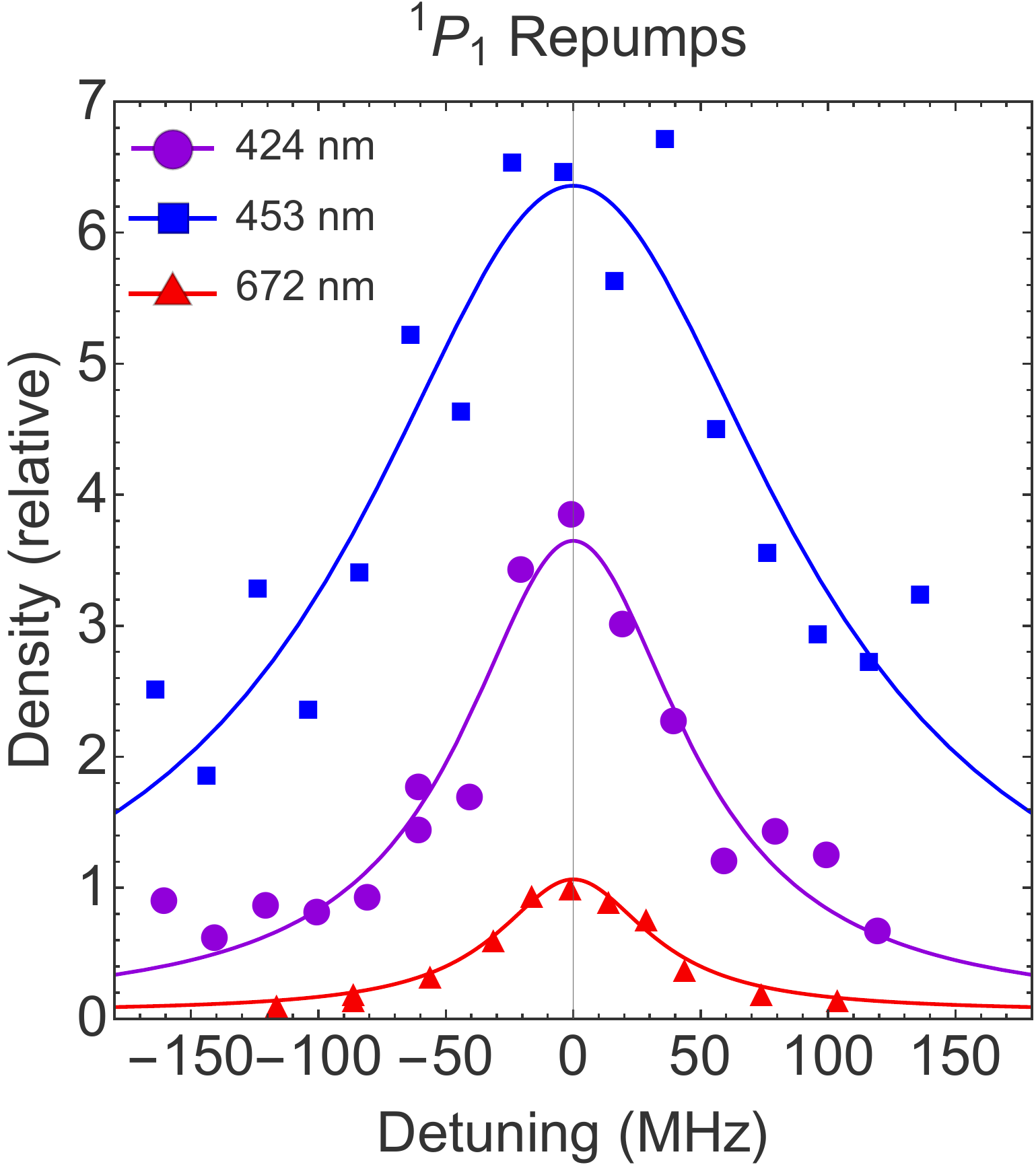}
\end{minipage}

\caption{\label{fig:densitydetuning}Measured calcium MOT density as a function of repumping laser detuning. Experimental data are shown by points, while Lorentzian fits are shown as lines. All measured densities are scaled to the peak MOT density achievable with the standard 672 nm repumping scheme.}
\end{figure}

\begin{figure}
\centering

\begin{minipage}[t]{\columnwidth}
\centering
\includegraphics[width=\columnwidth]{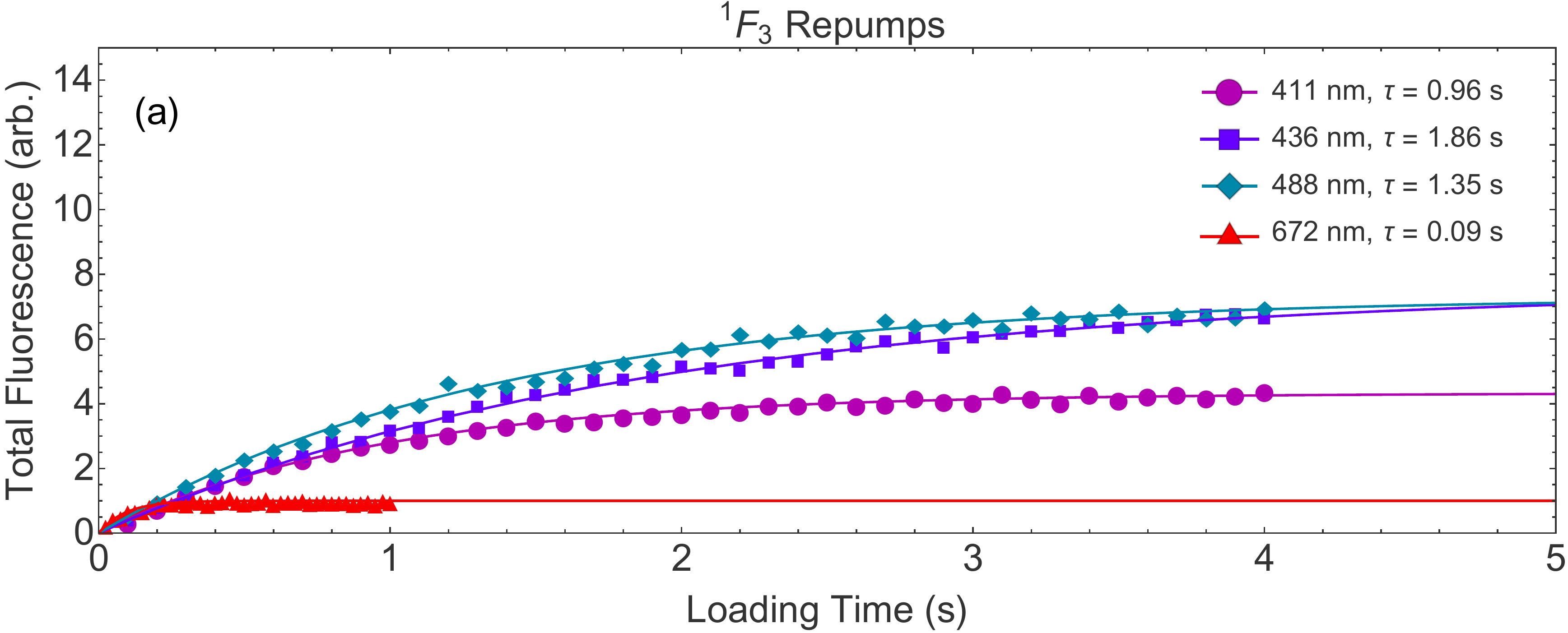}
\end{minipage}
\begin{minipage}[t]{\columnwidth}
\vspace{2pt}
\centering
\includegraphics[width=\columnwidth]{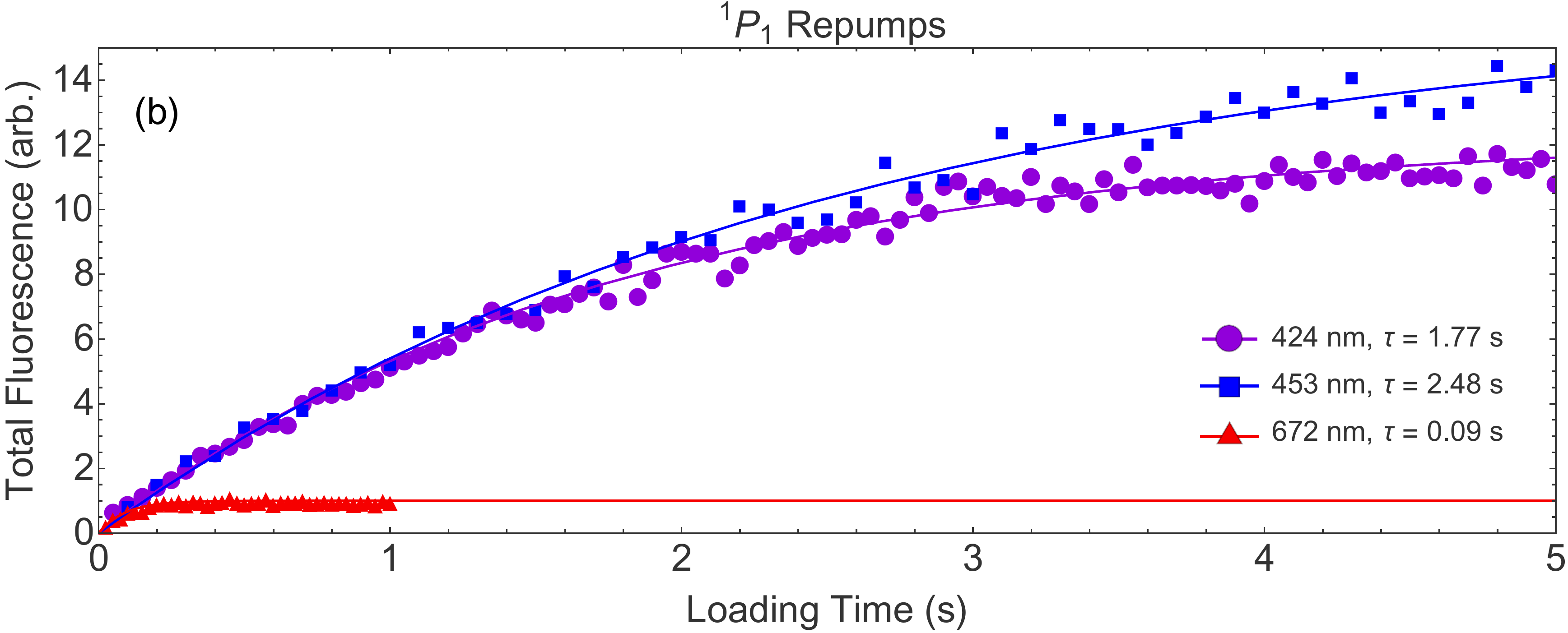}
\end{minipage}

\caption{\label{fig:lifetime}Measured Ca MOT loading curves for the (a) $^1F_3$ and (b) $^1P_1$ repump transitions, MOT fluorescence is plotted as a function of time elapsed after the cooling lasers are turned on; curves fitted to $N(t) = R\tau\left(1-e^{-t/\tau}\right)$ are shown alongside the data.}
\end{figure}

\begin{figure*}

\fbox{\includegraphics[width=.48\textwidth]{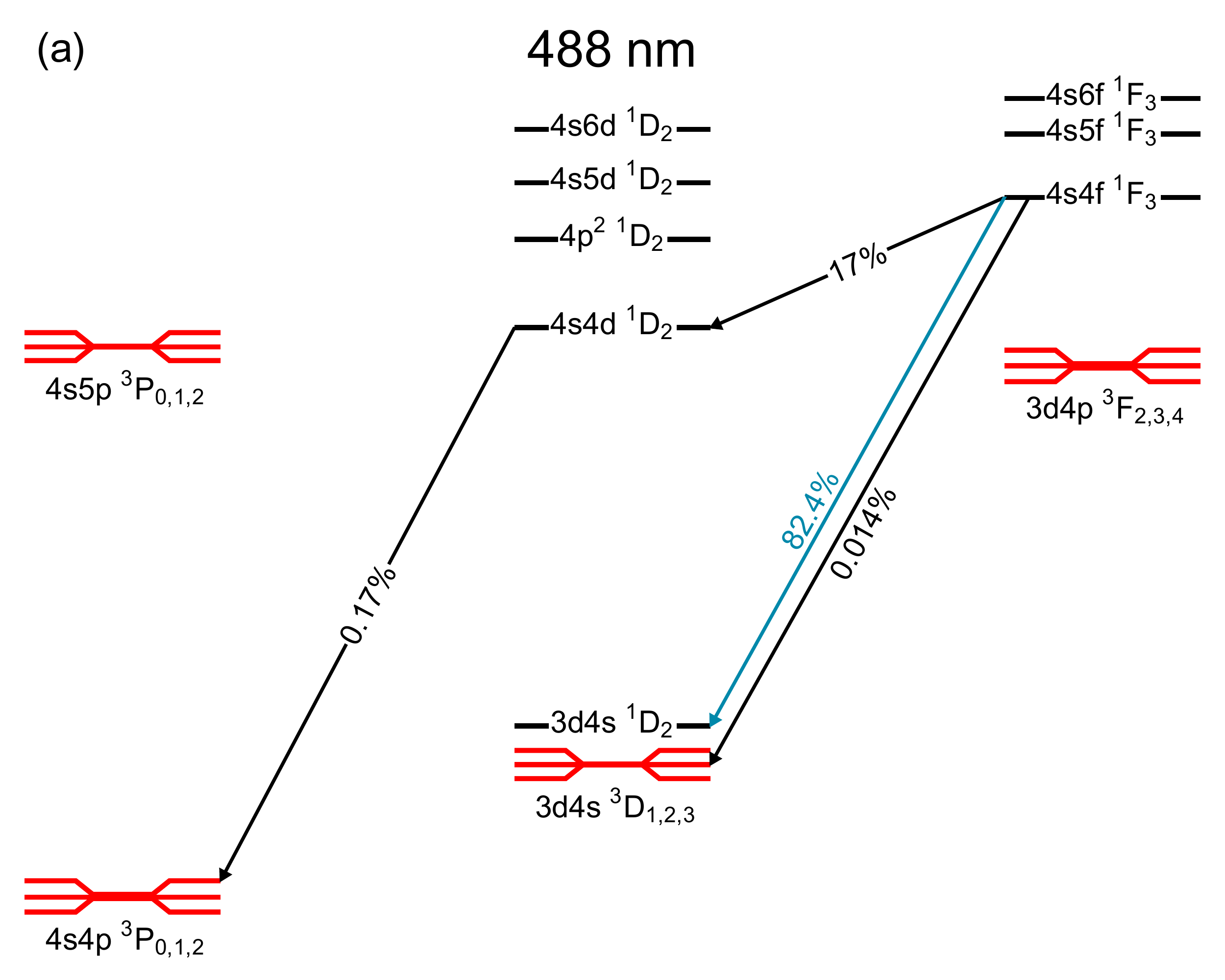}}
\fbox{\includegraphics[width=.48\textwidth]{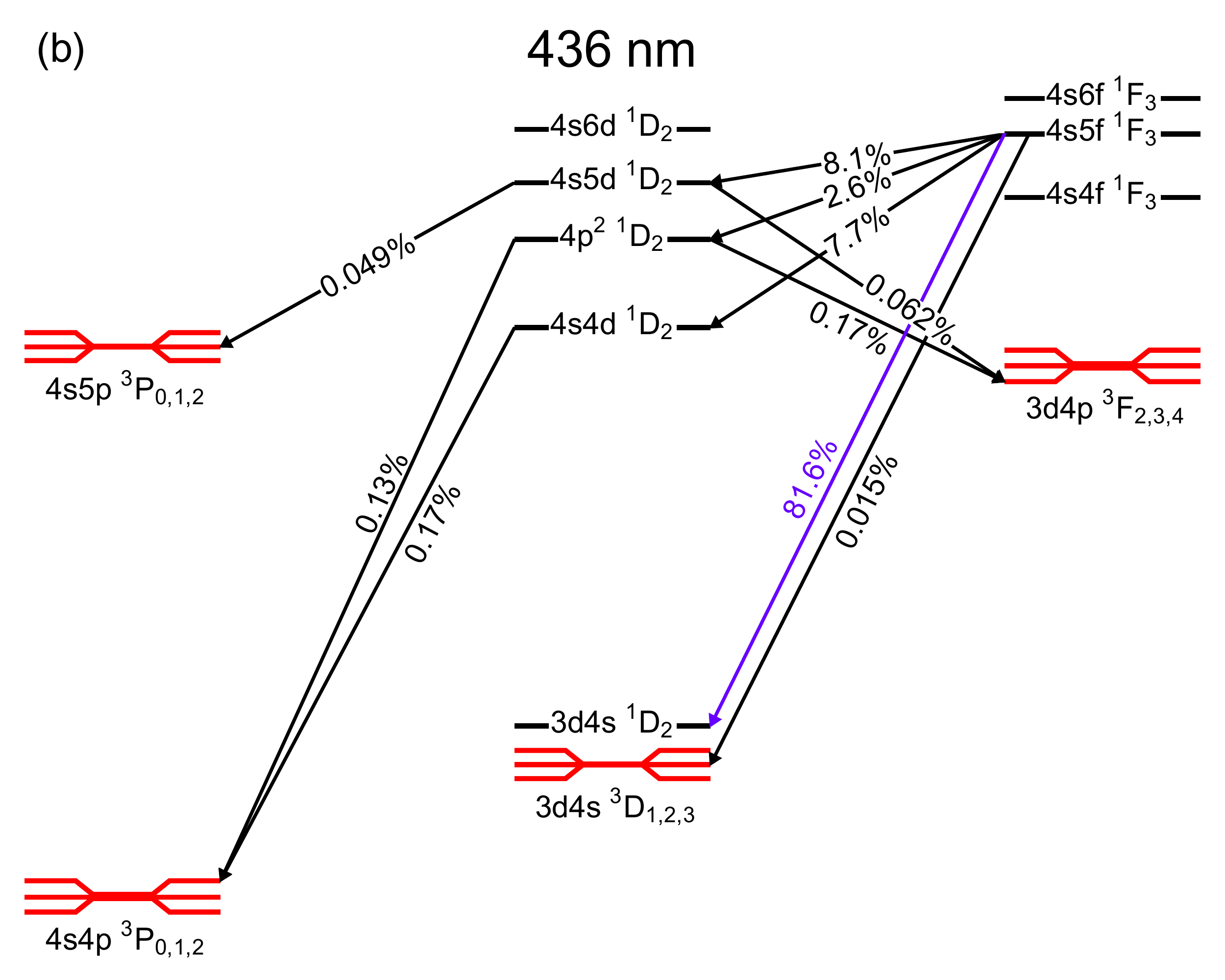}}

\fbox{\includegraphics[width=.48\textwidth]{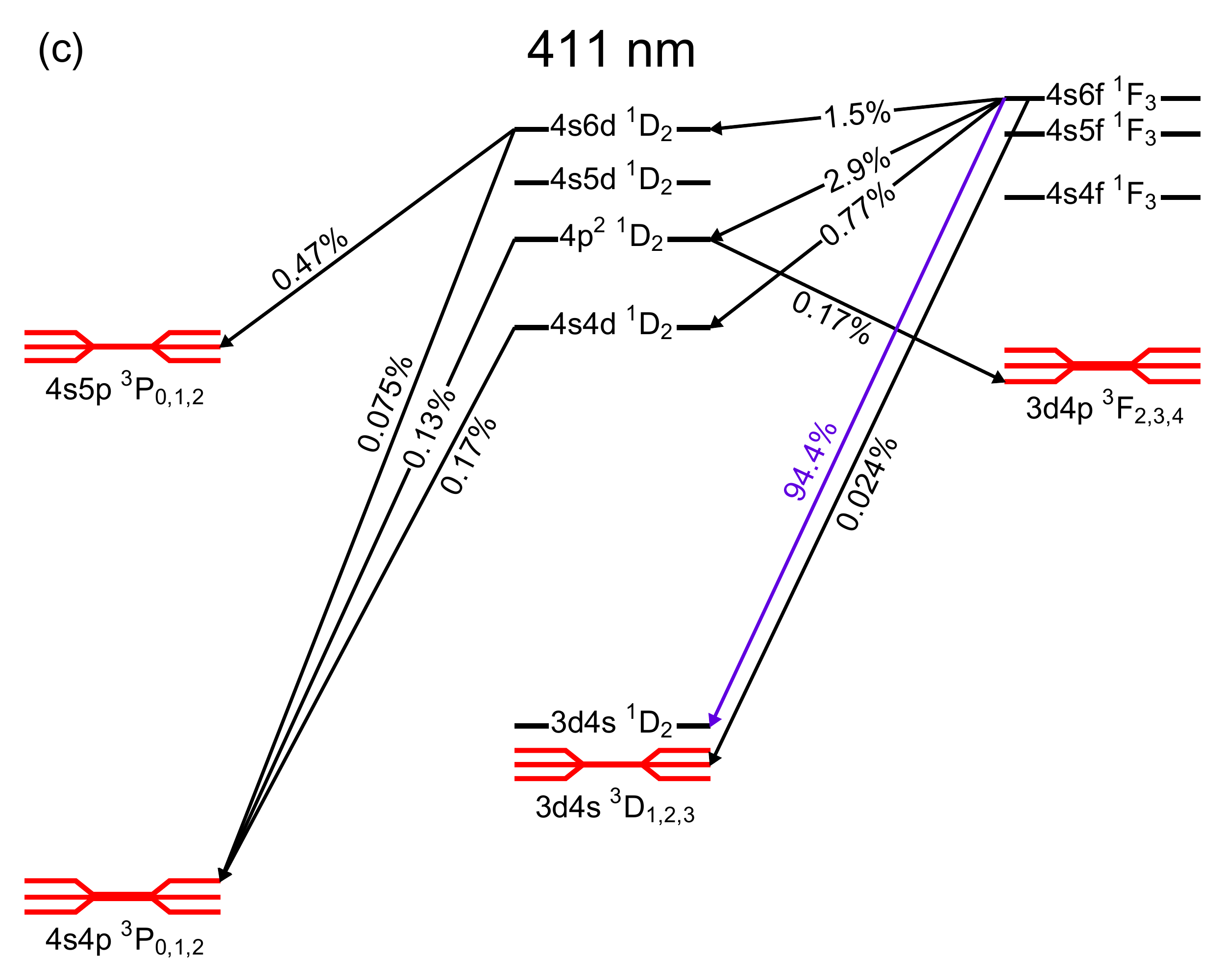}}
\fbox{\includegraphics[width=.48\textwidth]{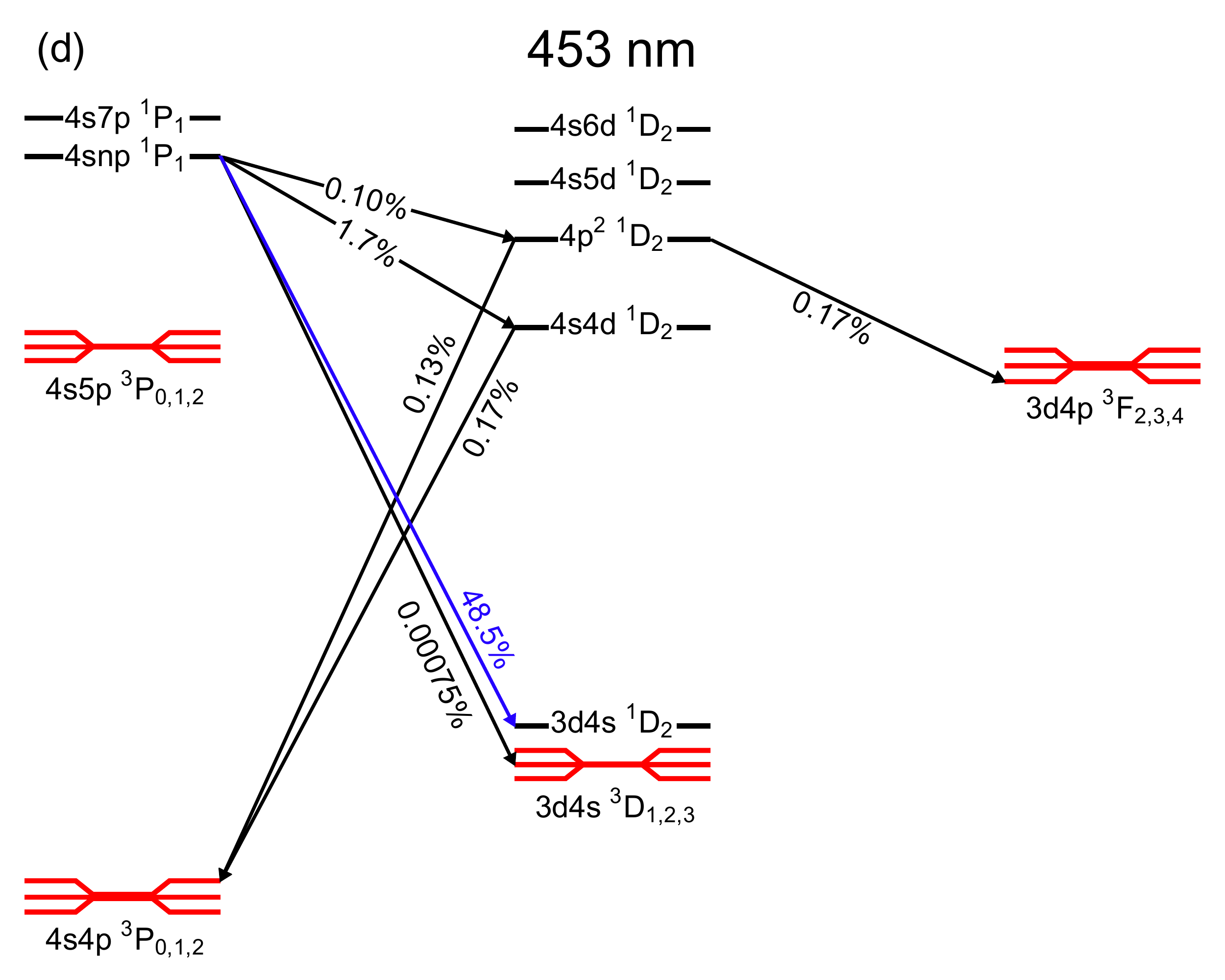}}

\floatbox[{\capbeside\thisfloatsetup{capbesideposition={right,center},capbesidewidth=.464\textwidth}}]{figure}[.50\textwidth]
{\caption{Simplified electronic energy level structures for the experimentally tested repumping schemes. $^1F_3$ repumps are shown in (a), (b), and (c), and $^1P_1$ repumps are shown in (d) and (e). Here we show only the most significant pathways into lossy triplet states, shown in red. Using only these branching ratios and the natural linewidths of the upper states, one can compare the approximate relative MOT lifetimes for each transition. This simple model reproduces the lifetime ordering of the more comprehensive 75-level rate equation model and also matches experimental results.}\label{fig:trip_decay_F_states}}
{\fbox{\includegraphics[width=.48\textwidth]{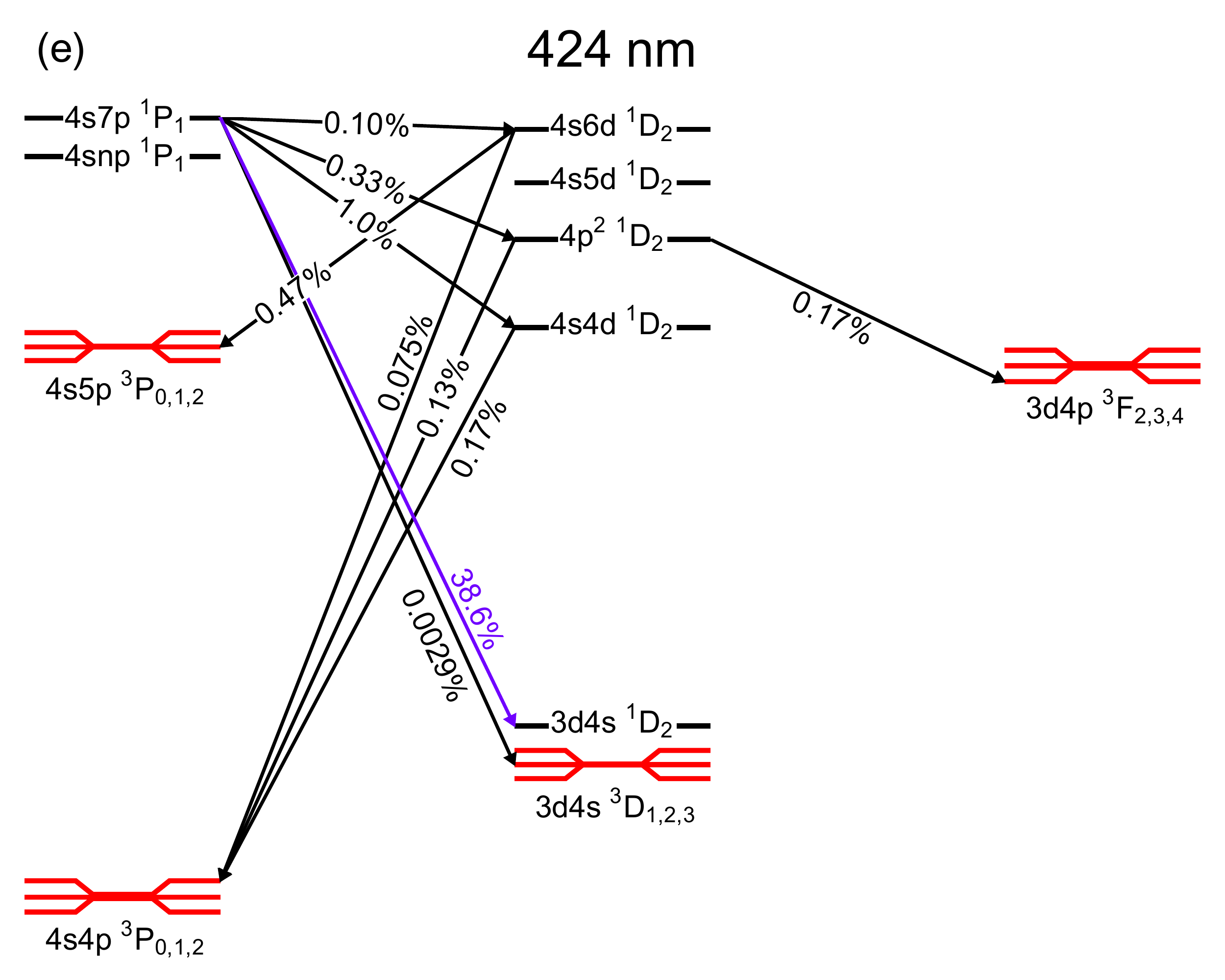}}}

\end{figure*}

For each single-beam repumping scheme, we characterize the MOT performance by measuring the MOT density, lifetime, and temperature. The density is measured using absorption imaging on the \pNine\ $\leftarrow$ \sOne\ transition. The MOT lifetime, $\tau$, is extracted by using fluorescence imaging to observe the number of trapped atoms, $N$, as the MOT is loaded from the oven at rate $R$ and fitting the data to the form $N(t) = R\tau\left(1-e^{-t/\tau}\right)$. The temperature, $T$, is found from the ballistic expansion of the Ca atoms after the MOT trapping beams are extinguished. For this measurement, the $e^{-1}$ waist of the cloud is extracted from absorption images taken after a variable time of expansion, and $T$ is extracted by fitting this data to the form $w(t>0) = \sqrt{w(t=0)^2 + \frac{2 k_B T t^2}{m}}$, where $k_B$ and $m$ are the Boltzmann constant and the mass of the Ca atom, respectively. The results of these measurements are shown in Fig.~\ref{fig:densitydetuning}-\ref{fig:lifetime} and Table~\ref{mainTable}. All of the experimentally explored alternative repumping schemes produce significantly denser MOTs at roughly the same temperature with longer optical pumping lifetimes.

Somewhat surprisingly, repumping to $^1F_3$ states leads to similar or sometimes better MOT performance than repumping to $^1P_1$ states. Population promoted to the $^1F_3$ states quickly decays to states with term $^1D_2$, which in turn primarily decay to the \pNine\ state. During this cascade, there is less decay into states of triplet character as compared to decays from some of the $^1P_1$ repumping states. Thus, despite the more complicated repumping pathway, repumping to the $^1F_3$ states can be very effective.

The relative performance of the $^1F_3$ repumping schemes can be explained by their branching pathways into lossy triplet states. The total MOT loss rate due to loss from an upper repump state is given by $\frac{d}{dt}N=-\Gamma_i f_{Loss} N_i$, where $N$ is the total number of atoms in the MOT, $N_i$ is the number of atoms in the upper repump state, $\Gamma_i$ is the natural linewidth of the upper repump state, and $f_{Loss}$ is the fraction of decays which lead to decay into the triplet states directly or indirectly. Of the three $^1F_3$ repump transitions experimentally tested, we approximate the relative values of $N_i$ by comparing the average number of repump transition cycles required before decay into another state. We use the calculated linewidths $\Gamma_i$ along with the most significant loss pathways to estimate $f_{Loss}$. 

Summarizing from Fig. \ref{fig:trip_decay_F_states}, the \fFortyTwo\ state decays with $\sim17\%$ branching into the $4s4d$ $^1D_2$ state, which has a branching of $\sim0.2\%$ into the \pFour\ state. The \fFiftyNine\ state decays to the $4s4d$ $^1D_2$, $4p^2$ $^1D_2$, and $4s5d$ $^1D_2$ states  with $\sim8\%$, $\sim3\%$, and $\sim8\%$ branching, respectively. The $4p^2$ $^1D_2$ state decays to triplet states with $\sim0.3\%$ branching, and the $4s5d$ $^1D_2$ state decays to triplet states with $\sim0.1\%$ branching. The \fSeventyFive\ state decays with branching ratio $\sim0.8\%$, $\sim3\%$, and $\sim1.5\%$ into the $4s4d$ $^1D_2$, $4p^2$ $^1D_2$, and $4s6d$ $^1D_2$ states respectively, the last of which decays with $\sim0.5\%$ branching into the $4s5p$ $^3P_1$ state. 
Using this method with only the branching ratios shown in Fig.~\ref{fig:trip_decay_F_states} and the natural linewidths of the upper repump states, we predict that the lifetime of the MOT $\tau_{488}$, $\tau_{436}$, $\tau_{411}$, using a 488 nm, 436 nm, or 411 nm repump should obey the relaton: $\tau_{436} > \tau_{488} > \tau_{411}$. This agrees with the observed MOT lifetimes. For the same reason, we expect repumping to the \fThirtyFour\ state with a 535 nm laser will exhibit poor performance. One can use this method to quickly estimate relative performances of potential repump transitions without developing a comprehensive rate model.

\begin{table*}[th]
\caption{\label{mainTable}Summary of the results of this work. Each row of this table lists the calculated and measured properties of an individual repumping scheme, with the most efficient repump transition to the \pFiftyThree\ state bolded. We attribute deviations between the rate model prediction for the MOT lifetime and the measured lifetime to inaccuracies in the calculated transition amplitudes. These inaccuracies are expected to be higher for the high-lying $F$-states, in agreement with the larger deviations seen between model and data for these states.}
\begin{ruledtabular}
\begin{tabular}{@c ^c ^c ^c ^c ^c ^c ^c ^c}
State & $\lambda$ (nm) & $f$ (THz) & $\rho_0$ (cm$^{-3}$) & N & $\tau$(s), model &  $\tau$ (s), exp. & T (mK) & Ca\textsuperscript{+} Production (relative)\\
\hline\\
\pNineteen\ & 672 &  446.150768(10) & 7.5$\times$10\textsuperscript{9} & 3.7$\times$10\textsuperscript{6} &0.098 &0.093(6)  & 4 & 1.0\\
\fThirtyFour\ & 535 & -- & -- & -- & 0.24 & -- & -- & --\\
\pThirtySeven\ & 504 & -- &-- & -- & 2.0 & -- & -- & --\\
\fFortyTwo\ & 488 & 614.393358(20) &2.1$\times$10\textsuperscript{10} & 2.7$\times$10\textsuperscript{7} & 0.76 &1.35(6) & 5 & 0.9\\
\rowstyle{\bfseries}
\pFiftyThree\ & 453 & 662.057094(20) &5.0$\times$10\textsuperscript{10} & 7.8$\times$10\textsuperscript{7} & 2.3 &2.48(8) & 5 & 0.8\\
\fFiftyNine\ & 436 & 688.180792(20) &2.8$\times$10\textsuperscript{10} & 2.8$\times$10\textsuperscript{7} & 0.94 &1.86(7) & 4 & 1.4\\
\pSixtyNine\ & 424 & 706.782952(4) &2.9$\times$10\textsuperscript{10} & 5.9$\times$10\textsuperscript{7} & 2.3 &1.77(6) & 5 & 1.7\\
\fSeventyFive\ & 411 & 729.478276(20) &2.5$\times$10\textsuperscript{10} & 1.6$\times$10\textsuperscript{7} & 0.48 &0.96(3) & 4 & 3.1\\
Ideal & -- & -- & -- & -- & 3.0 & -- & -- & --\\
\end{tabular}
\end{ruledtabular}
\end{table*}

Similarly, the MOT performance when repumping to the \pThirtySeven\ and \pSixtyNine\ states relative to the traditional \pNineteen\ state is understood by their primary branching ratios into triplet states. The \pThirtySeven\ state decays with $\sim0.006\%$ branching into the \dSix\ state, and the \pSixtyNine\ state decays with $\sim0.003\%$ branching into the \dSix\ state, while the \pNineteen\ state decays with $\sim0.9\%$ branching into the \dFive , \dSix , and $4s5s~^3S_1$ states.

Interestingly, the best MOT performance, in terms of number, density, and lifetime, is achieved by repumping to a highly configuration-mixed state, which we label as \pFiftyThree. Our calculations find this state is primarily composed of the mixture $4s7p$ (43\%), $4p3d$ (28\%), and $4s8p$ (13\%). 
The high performance of this repumping transition arises from two facts. First, its primary branching ratio to triplet states is $\sim0.001\%$ and the lowest of all repumping transitions explored here. Second, it exhibits a very high branching ratio of $\sim43\%$ directly back to ground \sOne\ state.

\begin{figure}[t]
\includegraphics[width=\columnwidth]{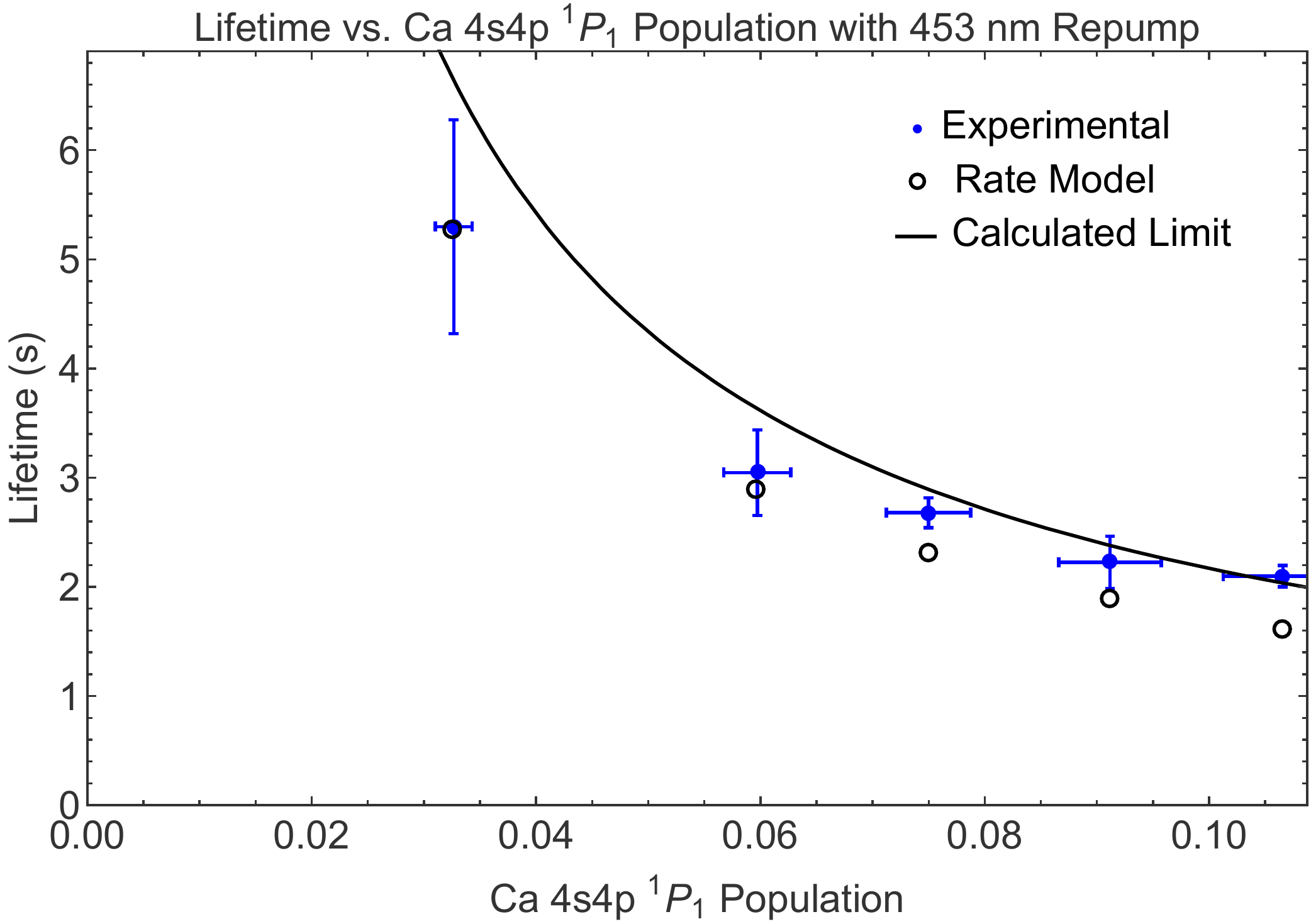}
\caption{\label{fig:lifetimeVSrhopp}Measured Ca MOT lifetime as a function of \pNine\ state population with a 453 nm repump. The measured lifetimes are shown alongside the rate model predictions and a curve representing the fundamental limit for any single repump laser scheme in a Ca MOT. This limit is the result of decay from the \pNine\ state indirectly to the \pTwo\ and $^3P_2$ states and is found as $0.24/\rho_{pp}$~s\textsuperscript{-1}.}
\end{figure}

Because the lifetime of the MOT when operating with the 453~nm repump ($\sim2.5$~s) is close to the idealized limit set by intercombination transitions from the \pNine\ state (3~s), we vary the intensity of the 423 nm cooling laser to measure the lifetime of the MOT as a function of the \pNine\ state population. Fig. \ref{fig:lifetimeVSrhopp} shows our results alongside the predicted lifetime from the rate model and the calculated limit of $0.24/\rho_{pp}$~s\textsuperscript{-1} set by the decay from the \pNine\ state indirectly to the lossy \pTwo\ and $^3P_2$ states -- here $\rho_{pp}$ is the population in state \pNine .  Our results show that the lifetime of the MOT in this scheme approaches this fundamental limit for any Ca MOT with a single repump out of the \dEight\ state. Therefore, repumping at 453~nm provides nearly the optimum performance for any imaginable single-repump scheme in Ca.

\section{C\MakeLowercase{a}\textsuperscript{+} production}
Due to its relatively light mass and high ionization potential Ca is especially useful in hybrid atom-ion traps as a sympathetic coolant~\cite{Rellergert2013}. However, as was recently identified~\cite{Sullivan2011, Schowalter2016}, Ca MOT operation can produce Ca$^+$ and Ca$_2^+$ through multi-photon and photo-associateve ionization, respectively. These ions then produce an unwanted heat load during the sympathetic cooling process. While techniques exist to cope with these nuisance ions~\cite{Schowalter2016}, it is advantageous to keep their production rate as low as possible. Therefore, we use time of flight mass spectrometry~\cite{Schowalter2012,Schneider2014} to measure the density normalized Ca$^+$ production rate for each of the tested repump lasers and compare to the Ca$^+$ production rate with a 672 nm repump. As shown in Table~\ref{mainTable}, we find that the largest Ca$^+$ production rate occurs with the 411~nm repump, a factor of $3.1\times$ compared to the Ca$^+$ production rate with the 672 nm repump. The 453~nm repump, which resulted in the MOT with the longest lifetime, highest density, and largest number of atoms also yields the lowest Ca$^+$ production rate.

\section{\label{Summary}Summary}
In summary, we propose seven alternatives to the traditional 672 nm repumping scheme for a Ca MOT and experimentally explore five of them. We find that all five produce significant improvements in MOT density and lifetime. Three of these repumping transitions appear particularly convenient from a technological perspective since they occur at wavelengths that are accessible by laser diodes, \textit{i.e.} 453~nm, 424~nm, and 411~nm -- with the middle transition of this list occurring at nearly the same wavelength as the cooling transition in Ca. The overall best MOT performance occurs for repumping at 453~nm on the \pFiftyThree\ $\leftarrow$ \dEight\ transition and results in a $\sim6\times$ and $\sim25\times$ improvement in density and lifetime, respectively, over the standard scheme. According to our rate model, this lifetime is near the maximum theoretical lifetime that can be achieved in a Ca MOT with a single repump laser from the \dEight\ state.

In all cases, the relative performance of the different repumping schemes can be understood by their branching into triplet states. Electronic population in these states typically ends up in either the \pTwo\ or $^3P_2$ state, which due to their long spontaneous emission lifetimes are lost from the MOT. For this reason, if a Ca MOT lifetime beyond that of $\sim5$~s is desired it would be necessary to add additional lasers to repump from the \pTwo\ and \pFour\ states as is done in Sr~\cite{Ludlow15}. However, even if these lasers are added, given the longer lifetime of the \pThree\ and \dFive\ states as compared to their analogues in Sr, it will likely be necessary to retain the 453~nm repump for optimal MOT operation.  

Finally, due to their similar atomic structure it may be possible to apply this repumping scheme in other Group 2(-like) atoms. For example, in Sr MOTs we speculate that repumping on the $5s8p$ $^1P_1 \leftarrow$ $4d5s$ $^1D_2$ transition at 448~nm may be beneficial since it would return population from the $4d5s$ $^1D_2$ more quickly than in the typically employed scheme and thereby increase the achievable optical force. A likely less efficient, but perhaps technologically simpler repumping pathway would be to drive the $5s6p$ $^1P_1 \leftarrow$ $4d5s$ $^1D_2$ transition at 717~nm. In both of these cases, however, it may be necessary to retain the lasers used to repump population from the $5s5p$ $^3P_0$ and $^3P_2$ states as the larger spin-orbit mixing in Sr increases the parasitic intercombination transitions from \textit{e.g.} the $5s5p$ $^1P_1$ state.

\acknowledgements
We would like to thank M. Kozlov for help with the CI-MBPT package~\cite{Kozlov2015CPC}.
This work was supported by the National Science Foundation (PHY-1205311 and PHY-1607396) and Army Research Office (W911NF-15-1-0121 and W911NF-14-1-0378) grants. YMY thanks support of the National Natural Science Foundation of China, Grant No. 91536106.

\bibliography{mybib,andrei}

\end{document}